\documentclass[3p, onecolumn,preprint,amsmath,amssymb]{elsarticle}
\usepackage{graphicx}
\usepackage[citecolor=blue,colorlinks=true,linkcolor=blue]{hyperref}
\usepackage{dcolumn}

\begin{document}

\title{Displacement cascades and defects annealing in tungsten, Part II: Object kinetic Monte Carlo Simulation of Tungsten Cascade Aging}

\author[rvt]{Giridhar Nandipati\corref{cor1}}
\ead{giridhar.nandipati@pnnl.gov} 
\author{Wahyu Setyawan}
\ead{wahyu.setyawan@pnnl.gov}
\author[rvt]{Howard L. Heinisch}
\author[rvt,kjr]{Kenneth J. Roche}	
\author[rvt]{Richard J. Kurtz}
\author[bdw]{Brian D Wirth }
\address[rvt]{Pacific Northwest National Lab, Richland, WA, 99352, USA}
\address[kjr]{Department of Physics, University of Washington, Seattle, WA 98195, USA}
\address[bdw]{University of Tennessee, Knoxville, TN 37996 USA}

\cortext[cor1]{Corresponding Author}
\date{\today}

\begin{abstract}
We describe the results of object kinetic Monte Carlo (OKMC) simulations of the annealing of primary cascade damage in bulk tungsten using a comprehensive database of cascades obtained from molecular dynamics \cite{PM2} as a function of primary knock-on atom (PKA) energy and direction, and temperatures of 300, 1025 and 2050 K. An increase in SIA clustering but decrease in vacancy clustering with temperature combined with disparate mobilities of SIAs versus vacancies causes an interesting temperature effect on cascade annealing, which is quite different from what one would expect. The annealing efficiency (ratio of number of defects after and before annealing) exhibits an inverse U-shape curve as a function of temperature. In addition, we will also describe the capabilities of our newly developed OKMC code; KSOME (kinetic simulations of microstructure evolution) used to carryout these simulations
\end{abstract}
\maketitle
\section{Introduction}

The operation of fusion power reactors will impose very demanding operating conditions on the plasma facing materials. They have to withstand severe and variable surface heat loads, damage from energetic neutrons and ions, and erosion of material by ions (or neutrals) during reactor operation. \cite{W1} Tungsten is considered the primary solid material choice for the main chamber and divertor components in future fusion reactors due to its high melting point, low sputtering coefficient, high thermal conductivity, low transmutation probability, low tritium retention and good mechanical strength. \cite{W1,W2,W3}  In a fusion environment bulk tungsten will be exposed to energetic neutrons escaping the plasma. The deuterium-tritium fusion reaction produces neutrons with a characteristic kinetic energy of 14.1 MeV. Collisions of 14.1 MeV neutrons with tungsten atoms cause radiation induced defects, transmutation and changes in microstructure, hence, degradation of its physical and mechanical properties with irradiation dose. \cite{W4}Therefore, it is important to develop predictive models of microstructure evolution for tungsten under fusion relevant conditions.

Collisions of 14.1 MeV neutrons with tungsten atoms produce tungsten primary-knock-on atoms (PKA) with various recoil energies. PKA atoms then lose their acquired energy through a sequence of collisions with other atoms, known as a displacement cascade. The whole process lasts for up to tens of picoseconds, and the debris left behind consists of point defects and clusters concentrated in a small volume, called primary damage.\cite{MD1} The actual number of defects created in the displacement cascade, and their size and spatial distribution in the solid will ultimately determine their effect on the irradiated microstructure,\cite{FMD2} as determined by the subsequent reaction and diffusion kinetics.

A common and long-standing question is how sensitive is the long-term evolution of the damage to the specific features of the primary damage defect distributions produced on picosecond time scales? \cite{KMC5} A random distribution of Frenkel pairs compared to the heterogeneous distribution of defects obtained from MD simulations will lead to very different damage evolution. \cite{KMC6, KMC7, KMC8} Therefore, an obvious step in understanding this problem is to study the annealing of individual cascades. Accordingly, in this article, we present results of single cascade annealing simulations in bcc tungsten from an extensive MD database of cascades with PKA energies from 10 keV to 100 keV at temperatures of 300, 1025 and 2050 K using the OKMC method. OKMC simulations are intended to be an extension of MD simulations in time-scale. Accordingly, the OKMC method is used to carry out annealing simulations at exactly the same temperature at which the cascades were generated with MD but using a much bigger simulation box. This paper constitutes Part II of our joint MD-OKMC cascade simulations. The results of the MD simulations are presented in a companion paper as Part I \cite{PM2}.

\section{KSOME: kinetic simulations of microstructural evolution}

KSOME is a flexible and computationally efficient lattice-based OKMC code to simulate the long-time scale evolution of a microstructure under irradiation. Patterned after the original FORTRAN code ALSOME, \cite{KMC2} KSOME is significantly more sophisticated, faster and flexible. Objects of interest include vacancies, self-interstitial atoms (SIA), interstitial impurities and clusters of these defects, produced during irradiation. KSOME keeps track of the locations (more precisely their center of mass) of these objects (defects) in time. Objects or defects are characterized by their type, size, shape, position, and orientation. Significant attention has been given to make KSOME as flexible as possible. While type and size are the most commonly used parameters to describe a defect, KSOME allows use of additional parameters, if available, such as defect shape and orientation. As there is no limit to the number of parameters that may be needed to describe a defect, one goal is to enable the use as many parameters as needed (or practical). For instance, KSOME differentiates between a $<$110$>$ and a $<$111$>$ dumbbell orientation of SIA, thereby allowing each to have different diffusion-reaction properties as well as transformation into one another. Orientation can also be used to represent 1D diffusion of SIAs along one of the four $<$111$>$ directions in case of a body-centered cubic (bcc) lattice. It can even differentiate between defects with different Burgers vectors and geometries, which again allows them to have different diffusion-reaction properties as well as transformation into one another. For example, it can differentiate between different geometries of vacancy clusters like voids, platelets and SFTs (face-centered cubic (fcc) only).  The present version of KSOME can handle any combination of point defect clusters in an fcc or bcc lattice. For example, if V (vacancy), I (SIA), H (interstitial hydrogen), and He (interstitial helium) are types of point defects present in a system, then KSOME can handle diffusion, emission, transformation and reaction events of those defects including clusters of all combinations (e.g. He$_m$I$_n$, He$_n$V$_m$, H$_m$He$_n$, H$_m$V$_n$, H$_m$I$_n$, H$_m$He$_n$I$_p$ and He$_m$H$_n$V$_p$).

KSOME is a flexible and computationally efficient lattice-based OKMC code to simulate long-time scale evolution of microstructure under irradiation. Patterned after the original FORTRAN code ALSOME, KSOME is significantly more sophisticated, faster and flexible. Accordingly objects of interest are vacancies, self-interstitial atoms (SIA), interstitial impurities and clusters of these defects, produced during irradiation. KSOME keeps track of locations (more precisely their center of masses) of these objects (defects) in time. Objects or defects are characterized by their type, size, shape, position, and orientation. With KSOME, lot of attention was given to make it as flexible as possible.  While type and size are most commonly used parameters to describe a defect, KSOME allows use of additional parameters if available like shape, orientation and so on. We note that there is no limit on the number of parameters that can be used to describe a defect. The idea is to use as many parameters as needed to capture the relevant atomistic details about a defect. Therefore KSOME, for example, can even differentiate between a [$110$] dumbbell and [$111$] dumbbell thereby allowing them to have different diffusion properties while letting them transform into one another.  In another example orientation can be used to represent 1D diffusion of SIAs along one of the four $<$111$>$ directions. The present version of KSOME can handle any combination of point defect clusters in fcc and bcc lattices. For example, if V (vacancy), I (SIA), H (interstitial hydrogen), and He (interstitial helium) are types of point defects present in a system, then KSOME can handle diffusion, emission, transformation and reaction events of those defects including their clusters of all combinations

In principle, the KSOME code allows the inclusion of extended defects such as dislocations and grain boundaries, which it presently treats as unsaturable sinks, such that any defect that is absorbed by a sink is removed from the simulation box. Free surfaces are treated as absorbing boundaries, which also act as sinks. In the present version of KSOME all defects are treated as spherical objects with an associated isotropic strain field. We assume that the medium is homogenous and continuous, and that the interaction radius of defects is isotropic. KSOME is designed such that all the necessary information regarding the properties of defects such as type, size, location, orientation, anisotropy, and their diffusion-reaction processes, along with simulation parameters regarding the simulation box size, temperature and boundary conditions, are provided via text-based input files. Correspondingly, as additional information is needed and available, KSOME can be used to carry out very detailed OKMC simulations with only minimal work required to modify the diffusion and reaction event databases.  As with all KMC approaches, much valuable information about the evolutionary path of a given system can often be obtained by determining its sensitivity to the deliberate variation of input parameter values ? which is easily accomplished with KSOME.

\section{Simulation Details}

Simulations were performed with a cubic block of tungsten atoms having a side of length 512 lattice units or 162 nm, with each axis parallel to a $<$100$>$ direction of the crystal. In tungsten, cascades are compact, and the volume of the simulation box chosen is large enough that the results were insensitive, especially the fraction of defects lost to recombination, to the box size. Each defect is allowed to hop to one of eight bcc nearest neighbor lattice sites along $\frac{a}{2}$ $<$111$>$, where a is the lattice constant. Initial SIA and vacancy defect distributions obtained from MD cascade damage simulations were placed in the center of the box. Absorbing boundary conditions were adopted in all three directions i.e. when a defect diffuses out of the box it is no longer tracked and is removed from the simulation. Such defects, also sometimes called ?freely migrating defects? (FMDs), are counted as escaped defects and contribute to long-range defect migration. Many major microstructural changes, such as the development of voids, dislocation networks and radiation-induced precipitation, that occur at temperatures where point defect mobility is significant, depend on long range defect migration, i.e., over distances large compared to typical primary damage dimensions of a few tens of nanometers. \cite{FMD2} These microstructural changes will produce macroscopic changes such as embrittlement and swelling.

The values of migration energies and pre-factors for diffusion and binding energies of defects used in our annealing simulations are taken from the {\it ab initio} calculations of Becquart {\it et al}. \cite{PM1}  and references therein. In the present model all SIA clusters are assumed to be $<$111$>$ loops, and the merging of two SIA clusters results immediately in the formation of a single larger $<$111$>$ loop. SIA clusters of all sizes move with the migration energy of 0.013 eV.\cite{PM3} SIA clusters larger than size five are constrained to 1D diffusion along one of four $<$111$>$ axes. SIA clusters up to size five are allowed to change their direction of 1D motion via rotation and thereby perform mixed 1D/3D motion. The activation barrier for changing the direction of their 1D motion from one $<$111$>$ direction to another is 0.38 eV. MD does not provide information on the orientation of SIA diffusion and since no bias was observed in the orientation of SIA clusters in the cascades obtained from MD, the direction of 1D motion was assigned randomly to the SIAs at the start of a simulation. We assume that all interstitial clusters are glissile (mobile) since there is no information available on infrequently occurring immobile SIA clusters \cite{IM1, IM2} in these cascades. Recently, in Fe it has been shown using a self-evolving atomistic KMC (SEAKMC) \cite{SEAKMC}method that even those clusters which appear to be sessile and stable will ultimately transform to glissile $<$111$>$ clusters (however, this transformation happens on the time-scale of seconds which is beyond regular MD time-scales). \cite{CDA3} The migration/diffusion rates of SIA clusters vary with cluster size (n) according to $\nu_0$n$^{-\alpha}$($\nu_0 = 6 \times 10^{12} s^{-1}, \alpha = 0.5$). We note that due to the low migration barrier of SIA clusters, the increase in their diffusion rates with temperature is very small compared to the decrease in their diffusion rates with increasing size. 

For a single vacancy, the activation barrier for diffusion is taken as 1.66 eV, and vacancy clusters larger than five are assumed to be inactive i.e. they neither diffuse nor emit, although they interact with other defects if they are within the range of interaction. All vacancies migrate in 3D, and their diffusion rates decrease with cluster size (n) according to n$_o(q^{-1})^{n-1}$ ($\nu_0 = 6 \times 10^{12} s^{-1}, q = 1000$). The vacancy (SIA) dissociation rate is given by $\Gamma_d = \nu_d $exp($(E_m + E_d )/k_BT$), where E$_d$ is the binding energy of a vacancy (SIA) to a vacancy (SIA) cluster, and Em is the migration energy of a single vacancy (SIA). Other required parameters like capture radii, binding energies of vacancies and SIAs were obtained from Ref. \cite{PM1}.

\section{Results and Discussion}
\subsection{General Details}

A database of 15-20 cascades at each of the PKA energies ranging from 10 keV to 100 keV was created using MD simulations. Detailed information about how the cascade database was created, as well as the details of defect production and other additional details about the cascades can be found in Ref. \cite{PM2}. From MD cascades, two sets of clustering data were generated. The first set (the default set) uses the third (NN3) and the fourth (NN4) nearest-neighbor distances to define the clusters of SIAs and vacancies, respectively. The second set of clustering data uses the second nearest neighbor (NN2) for both. Note that OKMC simulations using both sets of data were carried out, and we found that the results were similar in both cases. Irrespective of the criterion used to identify clusters, clustering of SIAs increases with PKA energy and temperature, while vacancy clustering increases with PKA energy but decreases significantly at the highest temperature of 2050 K. In tungsten, due to the very small change of SIA cluster diffusion rate with temperature, clustering of SIAs and vacancies seem to have a significant effect on how cascades anneal.

For each initial MD cascade of the primary damage state, 20 OKMC aging simulations were performed with different random seeds. The results were averaged over both different cascades from the MD database and from the OKMC simulations with random seeds. Thus each data point obtained was averaged over 300-400 runs. Simulations were carried out at temperatures of 300, 1025 and 2050 K for up to 10 ns. Note that 1025 K and 2050 K correspond to 0.25 and 0.5 of the melting temperature of tungsten, respectively. We have also carried out simulations longer than 10 ns, but due to the very fast interstitial diffusion rates we found that almost all of the recombination and coalescence events happen in the very first few nanoseconds of the simulation; therefore we show results up to only 10 ns for all temperatures and PKA energies. Beyond 10 ns we find the most frequent events are interstitial clusters escaping the simulation box. To understand the annealing of single cascades we have also examined runtime information such as defect densities, reaction events (e.g., recombination, coalescence, emission, and transformation), average cluster sizes and their size distributions, and number densities of defects escaping the simulation cell as a function of time. To compare annealing of cascades across different PKA energies and temperatures, the numbers of surviving defects were normalized by the average number of defects at the start of the simulation. This allows us to compare the numbers of defects lost to intracascade recombination across the different PKA energies and temperatures.

\begin{figure}
  \centering
  \includegraphics[width=3.0in]{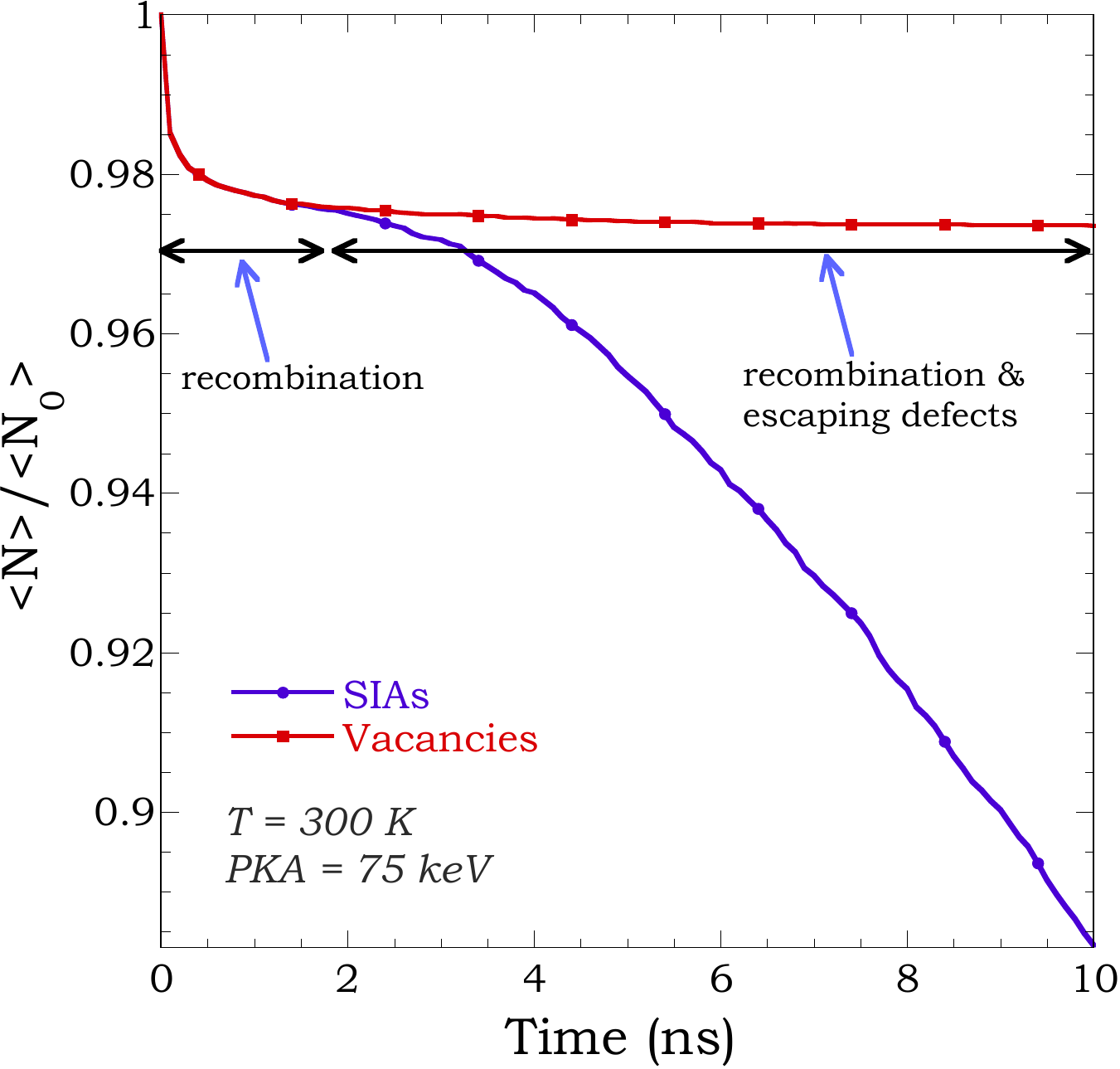}
  \caption{Surviving fraction of SIAs and vacancies as a function of time for a PKA of 75 keV and at 300K. }
  \label{fig1}
\end{figure}

Fig. \ref{fig1}  shows a plot of the surviving defect fraction of all (n$_d$ = $\sum$nS$_d$, n = number of defect clusters of size S$_d$) SIA and vacancy type defects as a function of time for a PKA energy of 75 keV at 300 K. This plot shows common features of defect evolution across the entire range of PKA energies and temperatures we have investigated.  Fig.\ref{fig1} shows that the initial drop of surviving defects occurs due to recombination only (surviving fraction of SIAs and vacancies are equal) and this occurs in the very first few ns of simulated time. Beyond the recombination phase, the fraction of surviving SIAs continually drops as they diffuse beyond the boundary of the simulation box resulting in the total number of interstitial defects is no longer equal to the number of vacancy type defects (see Fig.\ref{fig1}). At 300 K, interstitials migrate very fast and very quickly diffuse away from the primary damage region while vacancy migration rate is so low that they are effectively immobile during the time-scale of the simulation. Because of their low mobility, vacancy clusters can be considered immobile in these simulations up to 2050 K. As seen from Fig.\ref{fig1}, after the initial decrease due to recombination with SIAs, the surviving fraction of vacancies remains nearly constant. Therefore, for convenience, the surviving fractions of SIAs are shown in all of the remaining plots, with the surviving fractions of vacancies shown as needed. As mentioned previously, the emission rate is determined by the sum of the binding and migration energies and this sum is large for both SIA and vacancy clusters. Therefore, for the present simulations, emission is infrequent even at 2050K.

\subsection{Effects of PKA energy}

\begin{figure*}
  \centering
  \includegraphics[width=5.6in]{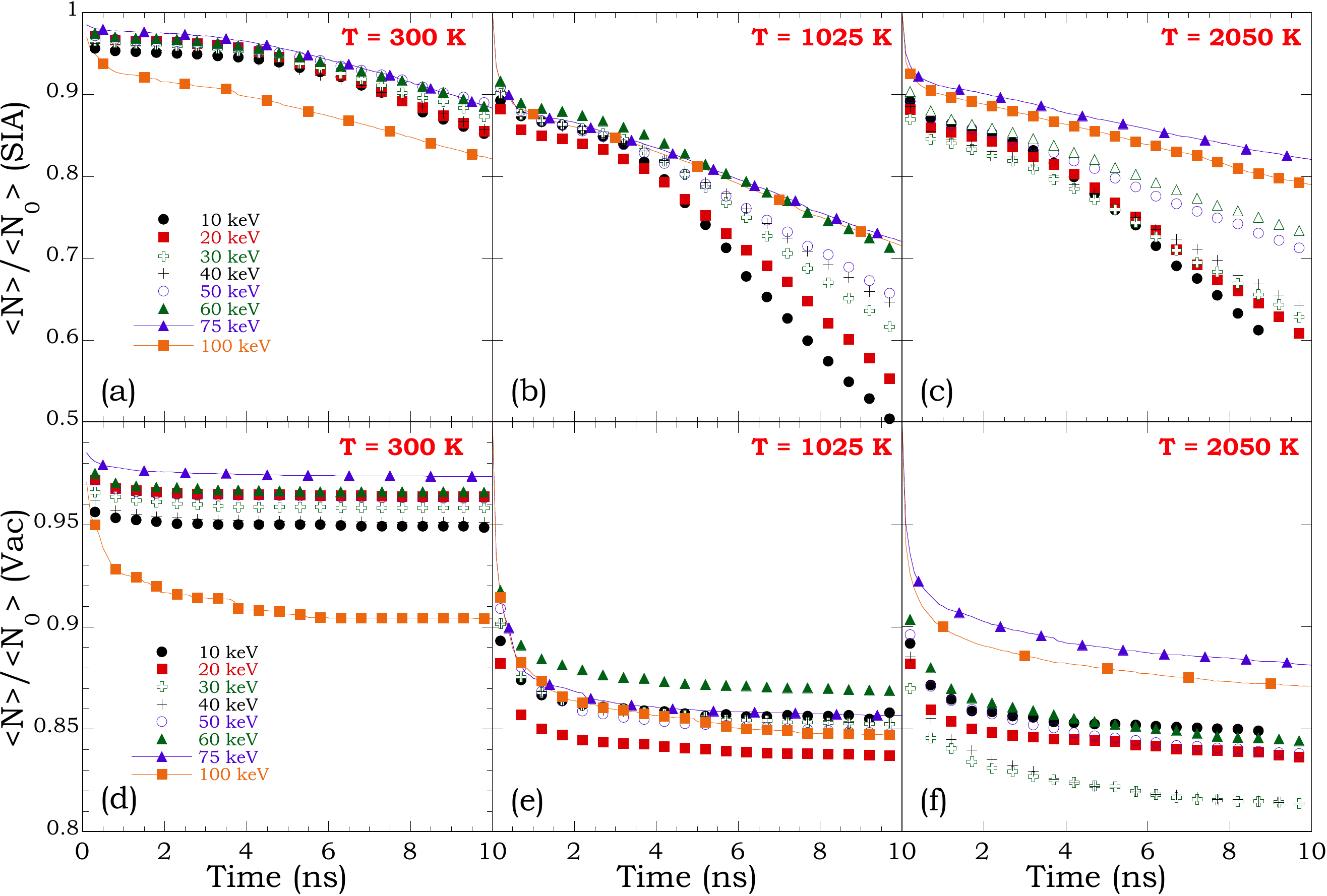}
  \caption{Surviving fraction of SIAs (a, b \& c) and vacancies (d, e \& f) as a function of time for different PKAs at 300, 1025 and 2050 K }
  \label{fig2}
\end{figure*}

Fig. \ref{fig2} shows the effect of PKA energy on the fraction of surviving defects at temperatures of 300, 1025 and 2050 K. For clarity the surviving fractions of SIAs and vacancies are plotted on different plots, but if they were plotted together they would look similar to Fig. \ref{fig1}. One can see from Figs. \ref{fig2} (d) \& (e) that at 300 K and 1025 K, after initial recombination the surviving fraction of vacancies plateaus. Within the 10 ns time-scale, due to their low mobility, vacancies do not migrate at 300 and 1025 K. Therefore recombination ceases once the SIAs migrate sufficiently far from the primary damage region, and consequently the number of surviving vacancies in the simulation box remains constant. Conversely, at 2050 K a slight decrease in the surviving fraction of vacancies still occurs from a few to 10 ns. This is primarily due to the migration of single vacancies. Since a single vacancy was not observed to diffuse to boundary of the simulation box within 10 ns, the decrease of vacancy concentration at 2050 K results from vacancy diffusion and recombination with, large SIA clusters that remain in the simulation box.

From Fig. \ref{fig2} it is also apparent that the fraction of defects lost to initial recombination tends to decrease with increasing PKA energy, especially at 300 and 1025 K. Perhaps this is due to the increased clustering of SIAs and vacancies with increasing PKA energy, which reduces the number of recombination events. Increased SIA clustering also reduces the fraction of SIAs that can perform rotation, thereby reducing recombination. We find that any fluctuation in the clustering fraction is also reflected in the fraction of defects lost to initial recombination. However this behavior is not consistent at 2050 K due to the almost non-existent vacancy clustering. Based on this reasoning the loss of defects due to recombination should be the lowest for the 100 keV cascades at 300 K. Interestingly, a bigger loss of defects to recombination is observed for 100 keV cascades (see Figs. \ref{fig2} (a) \& (d)) compared to cascades at other PKA energies, especially at 300 K. This can be explained based on the cascade morphology for 100 keV cascades at 300 K. A significant number of 100 keV cascades have very large vacancy clusters (>100) at their core and large SIA clusters (>60). The large drop observed in the average surviving fraction of both SIAs and vacancies is due to the annihilation of a large 1D diffusing SIA cluster, and the large vacancy cluster in the core of the cascade. At 1025 K, the reduced vacancy clustering with temperature does not create such a large effect, but the annihilation of large vacancy and SIA clusters does occur even at 1025 K.

After the initial loss of SIAs due to recombination, further loss of SIAs is primarily due to their escape from the simulation box. How SIAs escape the simulation box depends on the initial in-cascade SIA clustering and on the rate of SIA cluster growth. One expects that the number of Frenkel pairs and SIA clustering is higher for the higher PKA energies, resulting in a lower fraction of SIAs escaping the simulation box. By comparing how fast SIAs escape the box with SIA clustering from Fig.3 in Ref.\cite{PM2}, we find that the escape rate can be quite sensitive to in-cascade SIA clustering. For large PKA energies, even though SIAs escape at a lower rate, a very large fraction of them do escape, which contributes to long-range defect migration.

\subsection{Effect of Temperature}

\begin{figure*}
  \centering                          
  \includegraphics[width=5.65in]{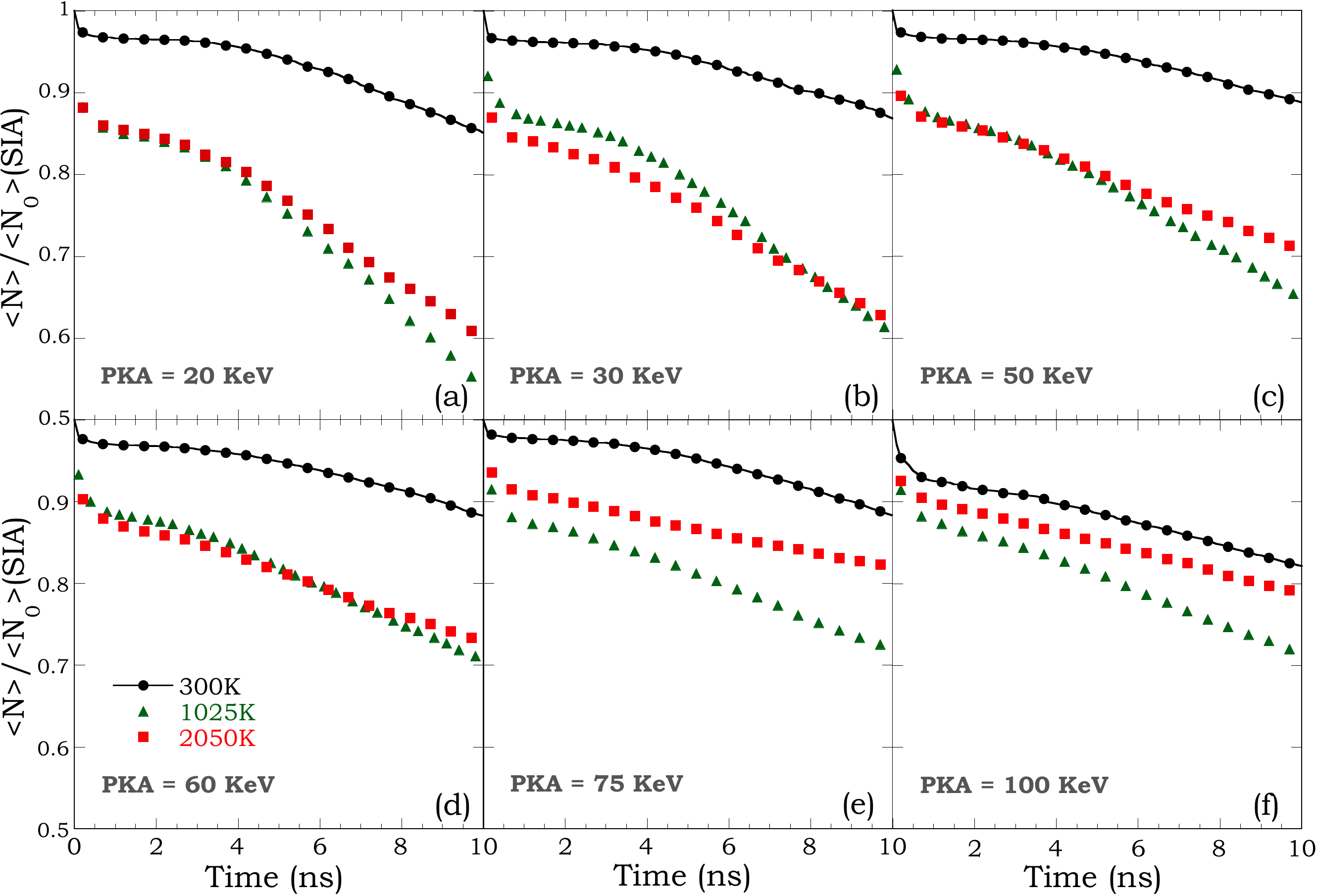}
  \caption{Surviving fraction of SIAs as a function of time at temperatures 300, 1025 and 2050 K  at PKAs of (a) 20 (b) 30 (c) 50 (d) 60 (e) 75 and (f) 100 keV}
  \label{fig3}
\end{figure*}

 It is expected that with increasing annealing temperature, more recombination events due to increased rotation of SIA clusters and faster migration of vacancies thereby increasing the loss of defects due to initial recombination. However, From Fig. \ref{fig3} shows that in tungsten this is not the case. From Fig. \ref{fig3} one can see that for cascades from PKA energies of 20 to 60 keV, the fraction of SIAs lost due to recombination appears to increase slightly with temperature. (see Figs. \ref{fig3}(a), (b), (c) \& (d)). However, in most cases, the fraction of SIAs lost to recombination is very similar at 1025 and 2050 K for PKA energies up to 60 keV. Whereas for cascades at 75 and 100 keV, the fraction of SIAs lost to recombination increases with temperature from 300 K to 1025 K, then decreases at 2050 K (see Figs. \ref{fig3}(e) \& (f)), i.e. it exhibits an inverse U-shape curve as a function of temperature. Note that the dotted line of filled red squares in Fig. \ref{fig3} represents the fraction of surviving SIAs at 2050 K, which falls between the observed response at 300 K and 1025 K for 75 and 100 keV cascades. Also, with increasing temperature one expects that the fraction of SIAs escaping the simulation box should increase with temperature. However, it is obvious from Fig. \ref{fig3} that at 2050 K, SIAs escape the simulation box at a slower rate (smaller slope) than at 1025 K for all PKA energies. Once again, even though SIAs escape at a lower rate, a very large fraction of them escape over time, contributing to long-range defect migration.
 
Among all the parameters that impact cascade aging, the short-time (100ns) the key variables are the initial spatial and size distribution of SIAs and vacancies, which depends on temperature and PKA energy, and the relative mobilities of those defects/defect clusters, which depends on cluster size and temperature. In tungsten, SIAs are highly mobile while the mono-vacancies have very slow mobility, and cascade-induced SIA clustering increases with temperature, but vacancy clustering decreases (see Fig. \ref{fig3} in Ref. \cite{PM2}). It appears that one of the main factors contributing to the temperature effect observed in cascade aging in tungsten is due to the temperature dependence of in-cascade SIA and vacancy clustering. From Fig. 5 in Ref. \cite{PM2} one can see that at 2050 K, vacancy clustering is infrequent (except for the very rare large cluster) and large numbers of vacancies exist as isolated and are dispersed over a larger region in space, while SIAs are clustered into just a couple (2 to 4) very large clusters. This increases the probability that a diffusing mono-vacancy will find another mono-vacancy to form an immobile cluster rather than annihilate at a mobile SIA cluster, thereby reducing recombination. However, the vacancy mobility relative to interstitial may not be fast enough even at 2050 K. The morphology of the cascades at 2050 K coupled with fast SIA and slow vacancy mobility reduces the fraction of defects that recombine.

\section{Conclusions}

Starting from a comprehensive cascade database obtained from MD simulations, OKMC simulations have been carried out to anneal single cascades in bulk tungsten as a function of PKA energy and temperature up to 10 ns using our newly developed code, KSOME. In tungsten, due to very fast SIA and very slow vacancy mobility, the fraction of defects lost to intracascade recombination is strongly affected by the spatial and size distribution of defects in cascades. This seems to be a key reason for a much lower fraction of defects lost to recombination at 2050 K compared to that at 1025 K, especially for higher energy PKAs. The rate at which SIAs escape from the simulation box appears to depend more sensitively on the SIA cluster size distribution than on PKA energy. Nevertheless, depending on the PKA energy and the temperature at which cascades were initiated, 85-95\% of interstitials eventually will escape intracascade recombination and contribute to long-range defect migration.

The annealing behavior of a single cascade is determined by outcome of two key properties (1) the spatial and cluster size distribution of defects and (2) the relative mobilities of those defects (kinetic parameters). This inter-dependence is non-linear and non-intuitive. Therefore, it may be difficult to predict the annealing behavior without actually carrying out a simulation. Due to uncertainty in the kinetic parameters of SIA clusters it may be also important to carry out sensitivity studies to determine the variability of the observed behavior in tungsten and may also help in understanding some aspect of cascade aging results. For example, in our simulations we have not considered the linear dependence of SIA diffusion rate with temperature \cite{PM3}. These studies will be the subject of future research.

\section*{Acknowledgments}
The work described in this article was performed at Pacific Northwest National Laboratory, which is operated by Battelle for the United States Department of Energy under Contract DE-AC06-76RL0-1830. This study has been supported by the U.S. Department of Energy, Office of Fusion Energy Sciences and Office of Advanced Scientific Computing Research through the SciDAC-3 program. All computations were performed on CARVER at National Energy Research Scientific Computing Center (NERSC).

\section*{References}

\bibliographystyle{elsarticle-num}
\bibliography{references}
 \end{document}